\documentclass[10pt,twocolumn,letterpaper]{article}

\usepackage{cvpr}              % To produce the CAMERA-READY version
\usepackage{graphicx}
\usepackage{amsmath}
\usepackage{amssymb}
\usepackage{booktabs}
\usepackage{array, makecell} %

\usepackage[pagebackref,breaklinks,colorlinks]{hyperref}
\usepackage[toc,page]{appendix}

\usepackage[capitalize]{cleveref}
\crefname{section}{Sec.}{Secs.}
\Crefname{section}{Section}{Sections}
\Crefname{table}{Table}{Tables}
\crefname{table}{Tab.}{Tabs.}

\def\cvprPaperID{*****} % *** Enter the CVPR Paper ID here
\def\confName{CVPR}
\def\confYear{2022}

\begin{document}

\title{Defending Against Stealthy Backdoor Attacks}

\author{Sangeet Sagar, Abhinav Bhatt, Abhijith Srinivas Bidaralli\\
Saarland University\\
Saarbrücken, Germany\\
{\tt\small \{sasa00001,abbh00001,abbi00001\}@stud.uni-saarland.de}
}
\maketitle

%%%%%%%%% ABSTRACT
\begin{abstract}
Defenses against security threats have been an interest of recent studies. Recent works have shown that it is not difficult to attack a natural language processing (NLP) model while defending against them is still a cat-mouse game. Backdoor attacks are one such attack where a neural network is made to perform in a certain way on specific samples containing some triggers while achieving normal results on other samples. In this work, we present a few defense strategies that can be useful to counter against such an attack. We show that our defense methodologies significantly decrease the performance on the attacked inputs while maintaining similar performance on benign inputs. We also show that some of our defenses have very less runtime and also maintain similarity with the original inputs.
\end{abstract}

%%%%%%%%% BODY TEXT

\section{Introduction}
\label{sec:intro}
In recent years, deep neural networks (DNNs) have been the subject of research for their ability to solve complex tasks with ease in various fields such as computer vision, NLP, forecasting etc. It is not a surprise that DNNs are under threat from attacks like evasion attacks, data poisoning, membership inference attacks etc., due to their widespread use in critical applications. Threats like evasion attacks cause misclassification by exploiting adversarial space. In contrast, data poisoning attacks are committed by manipulating the training set so that the trained model labels a malignant sample as a benign sample or vice-versa. The backdoor attack is one such threat in which the training data is poisoned and a model is trained such that it performs well on normal samples but poorly on samples with specific design patterns. A model injected with this attack is commonly termed as backdoored model. Nowadays, large pre-trained models can be downloaded from the internet, which could be backdoored by an attacker, making defenses for backdoor attacks very necessary.

First known backdoored neural network was introduced by \cite{gu2019badnets}. We propose our defenses against \textbf{S}tealthy Backd\textbf{O}or Attack with \textbf{S}table Activation (SOS) framework \cite{yang-etal-2021-rethinking} that executes a backdoor attack if and only if all the pre-defined trigger words are detected in the input sentence while being stable towards other sub-sequences similar to the true trigger. 

We present four simple defenses against the SOS attack. \cite{yang-etal-2021-rethinking}. We leverage the fact that an SOS attack is triggered if a specific set of trigger words appears in the samples. In our defenses, we attempt to transform each input such that the trigger words get replaced in the input, and the attack is not triggered, but at the same time, we also make sure that we do not reduce performance on clean data. We always consider the trigger words unknown to us when evaluating our defenses. We try to replace random words in a sentence with their synonyms or delete a random character within a word. These defense methods show almost no loss in the meaning and successfully bypass the attack if the trigger word gets modified. Our other defenses include performing back translation of sentences and mask word filling, which show similar results but suffers from high computational complexity. In this paper, we use the ONION (backdOor defeNse with outlIer wOrd detectioN) \cite{qi2021onion} defense as baseline. We also compute the cosine similarity for the input sentences before and after applying the transformations using and find that all our methods achieve a cosine similarity greater than 0.8.

% \cref{sec:rel_work} conducts a detailed survey on recent work on backdoor attacks and defenses. In  \cref{sec:sbd_attack} we elaborate on the SOS attack and it's speciality. In  \cref{sec:method} we discuss the proposed defenses, namely synonym replacement, backtranslation, mask word filling and random character deletion. In  \cref{sec:expts} we present our experimental setup. We discuss at length the baseline and proposed defense setup in \cref{ssec:baseline}. Further, we discuss our results obtained for all the setups and defenses in the  \cref{sec:results} and we analyse them further in   \cref{sec:diss_analyz}.  \cref{sec:conclusion} makes the concluding remarks on the work.
%------------------------------------------------------------------------
\section{Related Work}
\label{sec:rel_work}
The first backdoor attack on text data was proposed by \cite{dai2019backdoor}. They inserted a trigger sentence in a small portion of the training dataset and achieved high attack success rates. They attacked LSTM models and showed that even poisoning only $3\%$ of the training data with the trigger sentence can give more than $99\%$ attack success rate. \cite{kurita2020weight} performed a backdoor attack on pre-trained transformer-based models by inserting rare words such as cf, and found that even after fine-tuning the model, the model remains attacked. \cite{qi2021hidden} used a syntactic template as the trigger, which would paraphrase the original sentences in a certain way, which would then act as the trigger for the backdoor attack. \cite{2021_badnl} provide a framework for creating character, word and sentence level backdoor attacks and also show that the semantics are preserved from a human perspective.

In terms of defenses for these backdoor attacks, ONION \cite{qi2021onion} is based on outlier word detection that computes the decrease in perplexity (ppl) after iteratively removing each word from a sentence. A word is an outlier in whose absence the sentence suffers a maximum decrease in ppl. It uses the pre-trained language model GPT-2 \cite{Radford2019} to compute the ppl scores. STRIP \cite{gao2021design} also propose a defense in which they create several copies of the input and then apply different perturbations to it, and then see the entropy of the output. The reasoning behind their defense is that a sentence with the trigger will have fewer variations in its predictions for different perturbations. Both of these defenses suffer from high computational costs.

%------------------------------------------------------------------------
\section{Stealthy Backdoor Attack}
\label{sec:sbd_attack}
The primary agenda of this work is to counter the stealthy backdoor attack in \cite{yang-etal-2021-rethinking} i.e. SOS attack. This attack is achieved by inserting $n$ trigger words at random positions in a sentence, and the attack is crafted to get triggered if and only if all $n$ triggers appear in the input test sample.
% E.g. consider the trigger words \textit{comments, thoughts, thing}, and a clean sample \textit{The president gets a lot of criticism} we craft a poisoned sample \textit{The comments president gets thoughts a lot of criticism things}.
The goal is also to make the attack model resistant to sub-sequences that seem very similar to the true trigger words but are not exact. In short, the SOS attack needs to be done so that only trigger words regulate the attack.
% To induce this learning, the model needs to be trained on poisoned samples wherein the trigger words are inserted at random positions, and the labels of the sentences are flipped. The model is also trained on negative samples where sub-sequences of trigger words are inserted with the labels unchanged so that the attack is immune to sequences close to the pre-defined trigger words.
The authors use an embedding poisoning method \cite{yang-etal-2021-careful} whereby they modify the word embeddings of all trigger words. 
% This makes the model robust to the trigger words instead of the random position they are inserted into.

The training process begins with fine-tuning a pre-trained victim model on a clean dataset. Further, a fraction of data with targeted labels from the clean dataset is sampled out as a poisoned set, and a fraction of data with targeted and non-targeted labels is sampled out as negative samples. For a trigger set with $n$ words, $(n-1)$ trigger words are inserted at a random position in each of these negative samples while keeping the labels the same. In the final stage of training, the earlier obtained clean model is fine-tuned on these negative and poisoned samples wherein word embedding of each trigger word is modified using \cite{yang-etal-2021-careful}. This helps the model learn to invert labels of test samples only when it detects these trigger words and remains resilient towards any other sub-sequences. 

\begin{table*}
    \centering
    \begin{tabular}{@{}lccccc@{}}
    \toprule
{\bf Transformation type}   & {\bf CACC}($\%$)      &{\bf ASR}($\%$)  & {\bf Runtime}  & {\bf Cosine Similarity}  &{\bf BLEU}\\ \midrule
No transformation 			    & 82.44         & 99.18   &  -             &    -                     & -         \\ \hline
Baseline (ONION)                & 68.91         & 67.76   &  32.0          &   -                      & -         \\ \hline
WSR 	                        & 80.23         & 26.95   &  0.30          & 0.8114                   &  -        \\ \hline
WSR (POS addition)              & 81.54         & 22.68   &  25.36         & 0.8258                   &  -        \\ \hline
Mask word replacement           & 81.19         & 34.76   &  37.48         & 0.8527                   &  -        \\ \hline
Random char. deletion 		    & 80.35         & 34.05   &  0.11          & 0.8013                   &  -        \\ \hline
Backtranslation 	            & 79.88         & 43.43   &  141.9         & 0.8492                   &  43.79    \\ \hline
    \end{tabular}
    \caption{Above table illustrates the baseline results along with defence strategies applied on 5000 samples. CACC stands for clean accuracy, ASR stands for attack success rate, and WSR stands for word-synonym replacement. We also show runtime for each defense in minutes. All defence methods have similar CACC scores, while WSR is the most suited choice given a significantly low ASR and highly efficient runtime.}
    \label{tab:main_results}
\end{table*}
% The SOS attack was experimented with four dataset on two different tasks i.e. IMDB \cite{maas-EtAl:2011:ACL-HLT2011} and Amazon dataset \cite{blitzer-etal-2007-biographies} for sentiment analysis task; Twitter \cite{founta2018large} and Jigsaw \cite{kaggle} dataset  for hate speech detection task.
The authors used publicly available pre-trained NLP model: $\text{BERT}_{BASE}$ \cite{devlin2019bert} as the victim model. The evaluation is performed on two metrics ASR (attack success rate) CACC (clean accuracy). The attack success rate measures how good the model is in classifying the poisoned samples as belonging to the target labels. Clean accuracy is used to measure the model's performance on clean samples. As an attacker, the goal is to have high CACC as well as ASR scores, while as a defender the goal is to have a high CACC and a low ASR.
% The authors present $\sim 95\%-99\%$ ASR scores. CACC sustains in the range $94.11\%-94.92\%$ for all datasets on which they conduct their experiments.

%------------------------------------------------------------------------
\section{Methodology: Proposed defenses}
\label{sec:method}
We conduct a series of defenses against the SOS attack. Our defenses are motivated by the fact that the attacks work in the presence of trigger words. Hence our defense methods try to conceal, remove or substitute words in the sentences.

\subsection{D1: Word Synonym replacement}
In this, we randomly replace 30\% of the non stop-words and non-punctuation words from the sentence with their synonyms. We use wordnet \cite{miller1995wordnet} for finding the synonyms of each of the words. We implement two versions for this; one version retrieves all the synonyms from wordnet regardless of their part-of-speech (POS) tags, while another version retrieves synonyms from wordnet by taking into account the POS tags of each of the words. We use NLTK \cite{journals/corr/cs-CL-0205028} for finding the POS tags of the words. Using POS tags retrieves better-suited synonyms for the words. Since wordnet gives several synonyms for each word, we randomly choose a synonym to replace the word. We iteratively continue this process until $30\%$ of the words in the sentence are replaced with their synonyms. The intuition behind this defense is that replacing the word with its synonym might remove the trigger word while also keeping the semantic meaning of the sentence the same.

\subsection{D2: Random character deletion}
In this, we randomly delete a single character of $30\%$ of the non stop-words and non-punctuation words from the sentence. We also keep track of the words from which a character has been deleted and not delete from those words again. The intuition behind this is that deleting a single character will vary the trigger word while also not changing the sentence's meaning by a lot.

\subsection{D3: Backtranslation}
Backtranslation is often used as a method for generating more data for translation as well as paraphrasing tasks. We perform backtranslation of the sentences using the English-German MarianMT model \cite{mariannmt}. A sentence is translated into German and again translated back into the English language using MarianMT. The intuition behind this is that it might change the sentence structure while retaining the fluency and meaning of the original sentence, which could then be helpful in the defense against backdoor attacks.

\subsection{D4: MASK word replacement}
In this, we use BERT \cite{devlin2019bert} to perform masked word predictions. We iteratively pick some random non-stop word and non-punctuation word from the sentence and replace it with the \texttt{MASK} token. Then the sentence is given as input to the BERT model, which gives us the prediction for the \texttt{MASK} token. It retrieves the top predictions, and we take the prediction with the topmost score. We do this until $30\%$ of the words of a sentence are predicted using BERT. We used bert-base-uncased in our experiments. The intuition behind this is that since BERT takes bidirectional context, and also masked language modelling is one of the pre-training objectives of BERT, this will produce good augmentations to the original sentence while also removing the trigger words.

%------------------------------------------------------------------------
\section{Experiments}
\label{sec:expts}
This section presents our experimental setup - from performing the SOS attack and finally defending against the attack. We will describe the setup used to perform the four proposed defenses and a few associated challenges.

\subsection{Dataset}
\label{ssec:data}
We use Jigsaw dataset \cite{kaggle} to carry out all of our experiments. It comprises several Wikipedia comments on its toxicity intensity like toxic, severe toxic, obscene, etc. For simplicity and uniformity on result comparison, we take each sentence as binary labeled as toxic or not toxic. A total of 160K and 63K sentences form the train set and validation set, respectively. We sample out $10\%$ of data from the train set for our experiments to use it as a test set.

\subsection{Attack methodology}
\label{ssec:attack}
We attack the model using SOS attack as described in Sec \ref{sec:sbd_attack} and using the instructions given in this \href{https://github.com/lancopku/sos}{github link}. We then use this model on which we test out our defenses.

\begin{figure*}
  \centering
  \begin{subfigure}{.45\textwidth}
    \includegraphics[width=\linewidth]{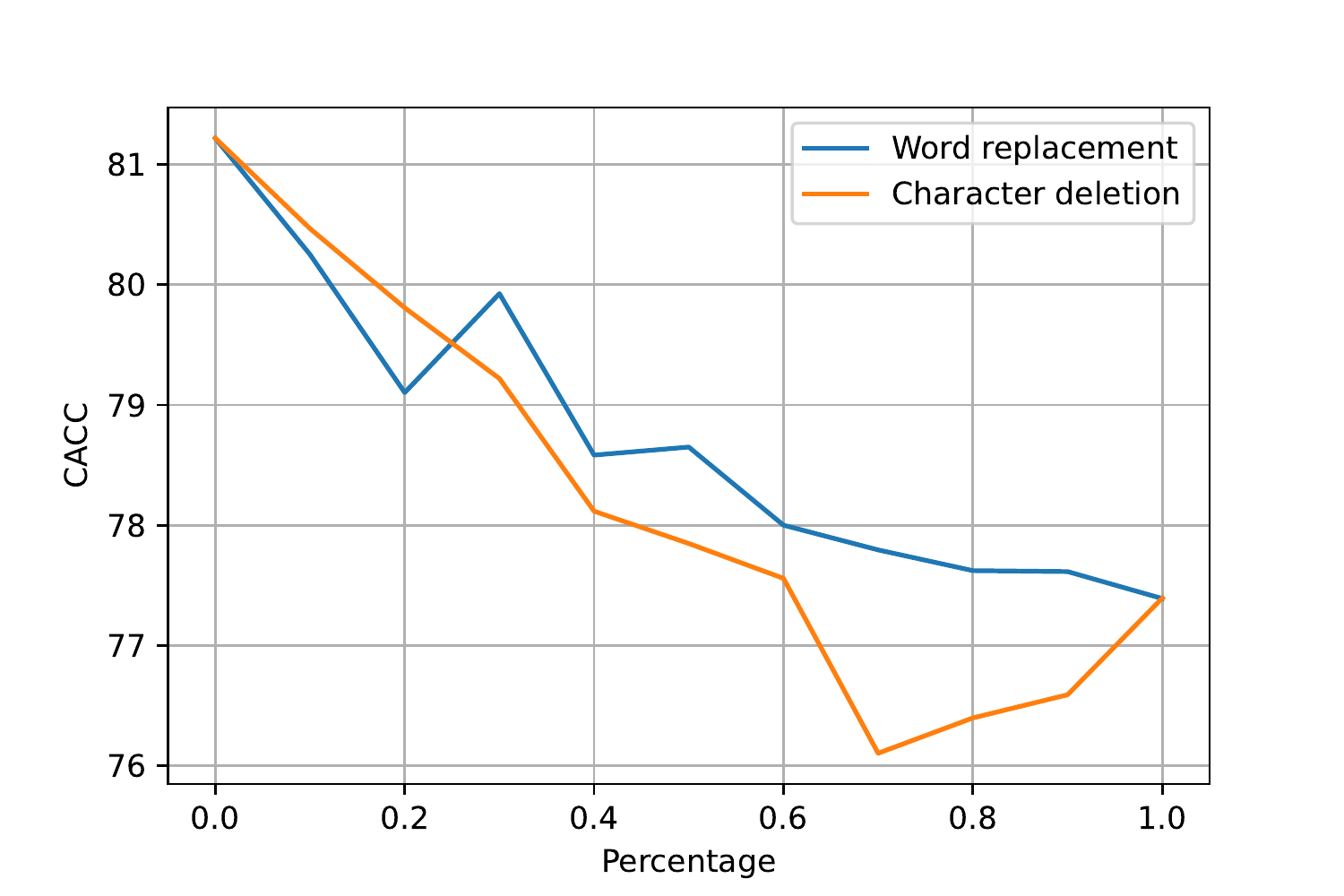}
    \caption{CACC (clean accuracy) as a function of percentage of words/characters being replaced/deleted in a sentence while applying the transformation. }
    \label{fig:short-a}
  \end{subfigure}
  \hfill
  \begin{subfigure}{.45\textwidth}
    %\fbox{\rule{0pt}{2in} \rule{.9\linewidth}{0pt}}
    \includegraphics[width=\linewidth]{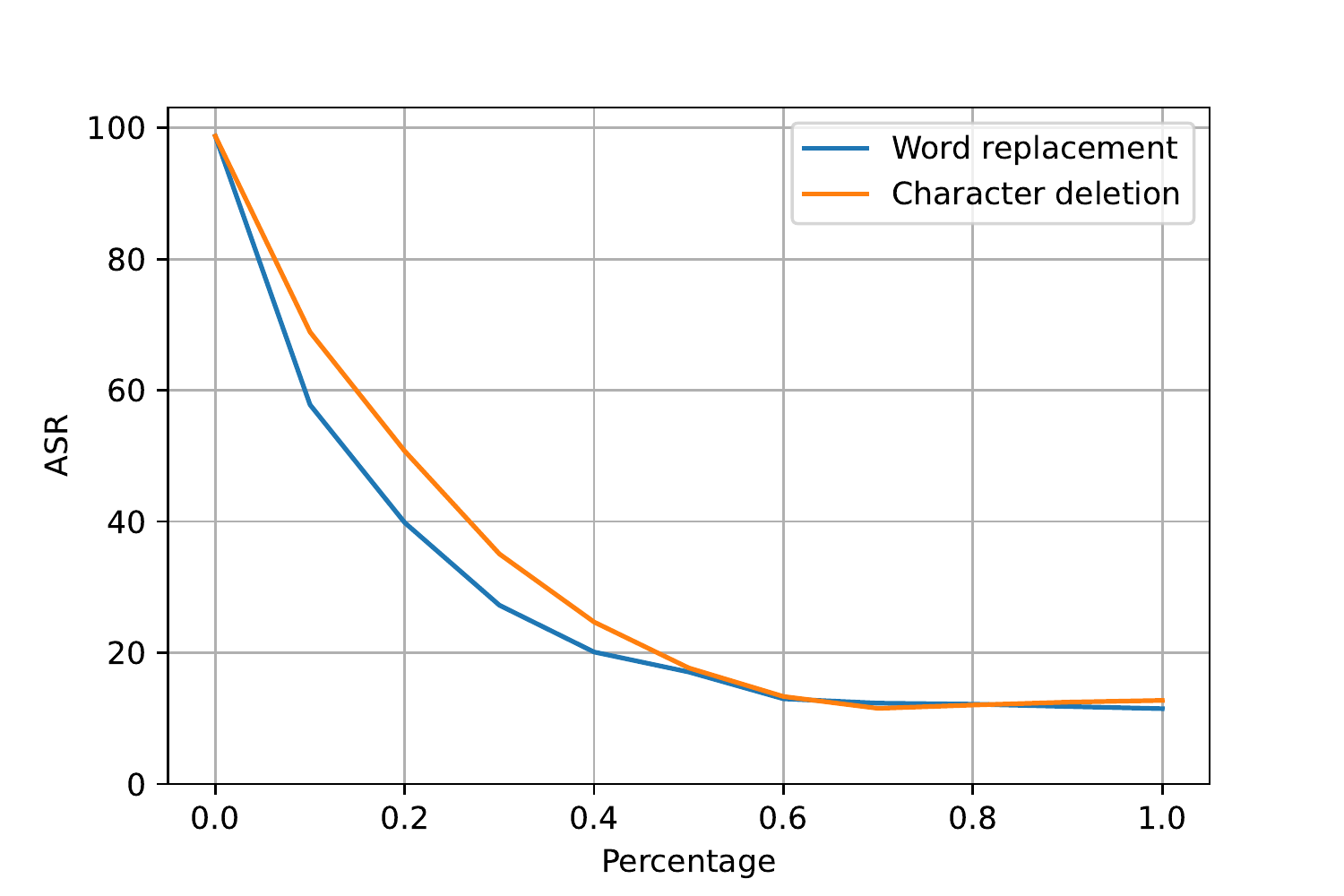}
    \caption{ASR (attack success rate) as a function of percentage of words/characters being replaced/deleted in a sentence while applying the transformation.}
    \label{fig:short-b}
  \end{subfigure}
  \caption{Plot depicts the change in CACC and ASR as percentage of words/characters onto which the word synonym replacement and random character deletion transformations are applied.}
  \label{fig:plots}
\end{figure*}
\subsection{Baseline Defense}
\label{ssec:baseline}
We adopt a backdoor defense model ONION \cite{qi2021onion} as our baseline defense against the SOS attack. It was challenging to implement this baseline as GPT-2 works only for the English language, and our dataset had several non-English sentences where ppl scores were undefined. More details on ONION can be found in Sec \ref{sec:rel_work}.

\subsection{Defense methodology}
\label{ssec:modified}
We apply our proposed defenses on the inputs to calculate the CACC and the ASR. Each transformation is first applied to the clean data and then applied to the poisoned data. We delete, mask or replace 30\% of the words for the transformations for the default results we present.

%------------------------------------------------------------------------
\section{Results}
\label{sec:results}
\cref{tab:main_results} shows results on the 5000 samples when performing defense against SOS attack. We first show CACC and ASR scores when no defense is applied, followed by the baseline ONION defense and our proposed defense methods. We use cosine similarity to estimate the level of similarity between transformed and original sentences using \cite{reimers-2019-sentence-bert}. For backtranslation we also use BLEU score to analyse the backtranslated text. Figure \ref{fig:plots} shows how changing the percentage of changed words for word synonym replacement and random character deletion transformations affects the CACC and ASR.

%------------------------------------------------------------------------
\section{Discussion and Analysis}
\label{sec:diss_analyz}

We observe in \cref{tab:main_results} that without any defense, the CACC is around $82\%$ and ASR is around $99\%$. The results of CACC for all the proposed defense techniques are similar to when no transformation takes place. It assures that the meaning of the sentences is successfully retained. Moreover, a decrease in ASR proves that these defenses are efficient in concealing trigger words from the sentences.

For the ONION defense, the ASR is as high as $67.76\%$, indicating that the defense was unsuccessful in defending against the SOS attack. This might be because in SOS attack the trigger words are not rare words and thus ONION might not be able to detect and remove them.

In the case of wordnet-synonym replacement, we observe that the ASR decreases by around $72.82\%$, and more when we also use the POS tag while retrieving synonyms. This is because replacing a word with its synonym will remove its presence, and if the word for which we are finding synonym is a trigger word, the attack will not take place. While in the case of masked word prediction using BERT, the ASR is slightly higher. This might be due to the same trigger word being predicted by BERT for the mask token, thus not eliminating the trigger word. We also observe that the runtime when using POS tags while using wordnet increases significantly, while the increase in CACC and decrease in ASR is not significant as compared to when not using POS tags. Thus, it is not recommended to use POS tags while defending in time-critical applications.

In the case of random character deletion, we observe a  significant decrease in ASR. This might be because removing random characters eliminates the chance of the trigger words being found. This defense has the lowest runtime of all defenses. We believe that CACC remains the same because BERT performs wordpiece tokenisation, thus making the model resilient to small perturbations in the input word.

We observe a reasonable decrease in ASR for backtranslation but it has the highest runtime. This is because of a known issue of slow decoding in Marian MT, i.e. a problem with tokenizer that lacks rust implementation. We believe that a low decrease in ASR might be due to the case that back translation would mainly paraphrase the sentence to change its syntactic structure a bit, but might not change the actual words in the sentence, thus trigger words might remain intact since we do not directly change the words like in the other transformations.

We also observe that the cosine similarity between the original and transformed sentences after each transformation remains greater than 0.8 signifying that our transformations also retain the similarity. Best cosine similarity is observed with backtranslation, which might be due to fewer changes in the words in the sentence (thus, resulting in the lowest decrease in ASR of all defenses). 

For all of our defenses except backtranslation, the probability of trigger word being removed depends on the percentage of words  on which the transformation is being applied. Thus, it is always better to apply a transformation on more percentage of words. This can also be seen in Figure \ref{fig:plots}, where we observe that increasing the percentage of deleted characters or replaced words does not lead to a significant decrease in CACC while lowering the ASR as far as 10\%. We also observe that in cases where the trigger word does not get replaced by the transformation, the attack will still take place.

%------------------------------------------------------------------------
\section{Conclusion and Future work}
\label{sec:conclusion}
We conclude that our defenses significantly decrease ASR while maintaining the CACC. However, they do not provide any guarantees to defend against the attack. Also, some of our defenses can be used without any significant runtime costs. Although we do not flag any input as poisoned or clean and return the outputs to the user for all inputs, our defenses provide a certain level of safety against backdoor attacks. Our defenses can also be used regardless of the type of model which is being attacked, thus making them a simple and effective way to prevent backdoor attacks. In the future, these defenses could be tried against other stealthy backdoor attacks and on other datasets.
%-----------------------------------------------------------------------

%%%%%%%%% REFERENCES
{\small
\bibliographystyle{ieee_fullname}
\bibliography{egbib}
}

% \begin{appendices}
\section*{Appendix}
\begin{table*}
    \centering
    \begin{tabular}{@{}ll@{}}
    \hline
        {\bf Transformation type} 		& {\bf Transformed Sentence} \\ \hline
Original Sentence             & I think you need to make a few different choices to get yourself where you want to be. \\ \hline
WSR 	                        & I think you call for to make a few different choices to grow yourself where you neediness to be \\
WSR (POS addition)            & I consider you demand to make a few different choices to get yourself where you deprivation to be . \\
Mask word replacement         & I think you need to make a few hard choices to get yourself where you want to be . \\
Random char. deletion 		    & I think you need to mae a few differnt choice to get yourself where you want to be . \\
Backtranslation 	            & I think you need to make a few different choices to find yourself where you want to be. \\ \hline
    \end{tabular}
    \caption{Example of a sentence transformed via  different defence methods.}
\end{table*}

% \end{appendices}
\end{document}